\documentclass{appolb}
\usepackage{graphicx}

\usepackage{amsmath}
\usepackage{amssymb}
\usepackage{amsfonts}
\usepackage{graphicx}
\usepackage{pstricks}
\usepackage{tikz}
\usepackage{mathtools}

\newcommand{\be}{\begin{equation}}
\newcommand{\ee}{\end{equation}}
\newcommand{\bqr}{\begin{eqnarray}}
\newcommand{\eqr}{\end{eqnarray}}

\begin{document}

\title{Advanced statistical methods to fit nuclear models}

\author{M. Shelley$^{a}$, P.Becker$^{a}$,  A.Gration$^{b}$ and A. Pastore$^{a}$
\address{$^{a}$ Department of Physics, University of York, Heslington, York, Y010 5DD, United Kingdom\\
$^{b}$ Department of Physics \& Astronomy, University of Leicester, University Road, Leicester LE1 7RH}
}
\maketitle


\begin{abstract}
We discuss advanced statistical methods to improve parameter estimation of nuclear models. In particular, using the Liquid Drop Model for nuclear binding energies, we show that the area around the global $\chi^2$ minimum can be efficiently identified using Gaussian Process Emulation. We also demonstrate how Markov-chain Monte-Carlo sampling is a valuable tool for visualising and analysing the associated multidimensional likelihood surface.
\end{abstract}
\PACS{  21.30.Fe       
             21.60.Jz        
             21.65.-f          
             21.65.Mn}
  

\section{Introduction}

In general,  nuclear models contain parameters that must be determined to best reproduce experimental results. These parameters are typically adjusted on data by means of a least-square procedure~\cite{bar89,kor10,dob14,bec17}, or more generally using Maximum Likelihood Estimator (MLE)~\cite{bar89}. 
MLE provides us essential information about the estimated parameters, such as errors and correlations.
We refer to Ref.~\cite{dob14} for a more detailed discussion.

With a MLE, there are two main challenges: how to find the maximum in  the parameter space, esepcially in case on mutlimodal likelihood surfaces~\cite{sil81} and how to provide realistic estimates of errors.

In the present article, we discuss the benefit of using the statistical modelling method Gaussian Process Emulation (GPE)~\cite{ohaganCurveFittingOptimal1978,sacksDesignAnalysisComputer1989,rasmussenGaussianProcessesMachine2006} and the Expected Improvement (EI) criterion~\cite{jonesEfficientGlobalOptimization1998,gra18} to explore the likelihood surface~\cite{bar89,kor10,gao13} and identify  the  location of the maximum in parameter space. 
GPE has already been adopted in several scientific domains to facilitate the use of computationally expensive models~\cite{ken06,gib12,sal14,gra18}. It has also been recently applied to nuclear physics~\cite{mcd15,pas17,neufcourtBayesianApproachModelbased2018}. In this paper, we combine GPE with the EI criterion to estimate the parameters of a  Liquid Drop Model  (LDM)~\cite{krane1988introductory}.
Through Markov-chain Monte-Carlo (MCMC) sampling, we explain how to visualise the multidimensional likelihood surface and extract the covariance matrix without calculating explicitly the Hessian matrix~\cite{dob14}.

The article is organised as follows: in section \ref{sec:gpe_ei}, we introduce the basic formalism of GPE statistical method. In section \ref{sec:ldm} we present the LDM and  we demonstrate in section \ref{sec:results} how GPE and EI can be used to iteratively find the maximum of the unimodal likelihood of the LDM. We finally provide our conclusions in section \ref{sec:conc}.



\section{Gaussian Process Emulation and Expected Improvement}\label{sec:gpe_ei}
Gaussian Process Emulation (GPE) is a regression method  that performs a smooth interpolation of the points $y_s$ of a data set,  providing associated confidence intervals. In a standard fitting procedure, one assumes a fixed relation between the independent variable $\mathbf{x}$ and the points $y_s=f(\mathbf{x}|\mathbf{p})$, where $\mathbf{p}$ represents a set of adjustable parameters.
It is not the case with GPE, where we treat the  points $y_s$ as a draw from a Gaussian process (GP), $\mathbf{y}$ with mean $0$ and covariance $R(\textbf{x},\textbf{x'})$. See Ref.~\cite{gra18} for a more general discussion.

We still need to make some assumption about the structure of the data. Such an assumption is done for GPE on the structure of the covariance matrix. In the present article, we assume that the latter is filled with the \emph{squared exponential} covariance function, so that each matrix element corresponds to
\begin{equation}\label{kernel}
R(\textbf{x},\textbf{x'})=\sigma^2_{k}\prod_{i=1}^d \mbox{exp}\Big(-\frac{(\textbf{x}-\textbf{x'})^2}{2\theta_i^2}\Big),
\end{equation}

\noindent where d is to the number of parameters and $\textbf{x},\textbf{x'}$ are the data points.
\newline \noindent We choose to emulate normalised training data, to avoid numerical issues that can arise during emulation. We therefore set the maximum covariance to be $\sigma^2_{k}=1$. The \textit{hyperparameters} $\theta_i$, often referred to as correlation lengths or length-scales, describe the level of smoothness of the emulated surface along each dimension of the parameter-space. 
These \textit{hyperparameters} are adjusted on training data using a MLE method. We refer to Ref.\cite{rasmussenGaussianProcessesMachine2006} for more details.
Given the particular Gaussian form of the covariance matrix in Eq.\ref{kernel}, we observe that neighbouring points will be strongly correlated, while points that are far from each other will be uncorrelated. The length-scale controls how fast such a correlation drops moving away from a given point, and therefore moving away from the diagonal of the covariance matrix.

\noindent GPE prediction at an unknown location $\textbf{x}^*$ is
\begin{equation}\label{gpe_mean}
\hat{\mu}(\textbf{x}^*)=r^T(\textbf{x}^*)R^{-1}  \mathbf{y},
\end{equation}
\noindent where $r(\textbf{x}^*)=R(\textbf{x},\textbf{x}^*)$ corresponds to the vector filled with the correlation of  $\textbf{x}^*$ with all the data points. GPE also easily provides the  $1\sigma$ confidence intervals as
\begin{equation}\label{gpe_var}
\hat{\sigma}(\textbf{x}^*)=R(\textbf{x}^*,\textbf{x}^*)-r^T(\textbf{x})R^{-1}r(\textbf{x}).
\end{equation}

\noindent These formulas ignore possible uncertainties on the points $y_s$ since we assume here that they come from a deterministic computer simulation. It is however possible to take them into account into GPE procedure~\cite{and12}.
In the following section, we introduce a model that will be used to illustrate how GPE practically works.

\section{Liquid Drop Model}\label{sec:ldm}

The LDM is a simple model that links the binding energy $B_{th}$ of a given nucleus to its number of neutrons $N$ and protons $Z$.
It is composed of five adjustable parameters $\mathbf{p}=\{a_v,a_s,a_c,a_a,a_p\}$ that are related to bulk properties of the nucleus and  defined as~\cite{krane1988introductory}
\begin{eqnarray}\label{bene}
B_{th}(N,Z)=a_v A-a_sA^{2/3}-a_c\frac{Z(Z-1)}{A^{1/3}}-a_a\frac{(N-Z)^2}{A}-a_p\delta(N,Z),
\end{eqnarray}

\noindent where $A=N+Z$, and 
\begin{equation}\label{eqn:pairing}
\delta(N,Z)=
\begin{cases}
+A^{-1/2} & \text{$Z,N$ even\;,} \\
0 & \text{$A$ odd\,,}\\
-A^{-1/2} & \text{$Z,N$ odd\,.}
\end{cases}
\end{equation}

\noindent To determine the optimal set of parameters $\mathbf{p}_0$, we construct the penalty function 
\begin{eqnarray}\label{chi2}
\chi^2=\sum_{N,Z\in\text{data}}\frac{(B_{exp}(N,Z)-B_{th}(N,Z))^2}{\sigma^2(N,Z)}\;
\end{eqnarray}

\noindent to be minimised. The experimental energies, $B_{exp}(N,Z)$, are extracted from Ref.~\cite{wang2012ame2012}. Since Eq.\ref{bene} is not suitable for describing light systems, we exclude all nuclei with $A<16$. We also choose  not to take into account those with experimental errors on binding energies larger than 100 keV.
This LDM model has a large variance of a few MeV on average ~\cite{bertsch2017estimating,pas18b} making the residuals larger than any experimental errors. The weights in Eq.\ref{chi2} are therefore unimportant and we have fixed them as $\sigma^2(N,Z)=1$ for all nuclei. We also recall for the sake of clarity that the $\chi^2$ function is directly linked to the likelihood function as $\mathcal{L}\propto e^{-\chi^2}$.

\section{Results}\label{sec:results}

\begin{figure}[h!]
\centerline{%
\includegraphics[width=0.80\textwidth]{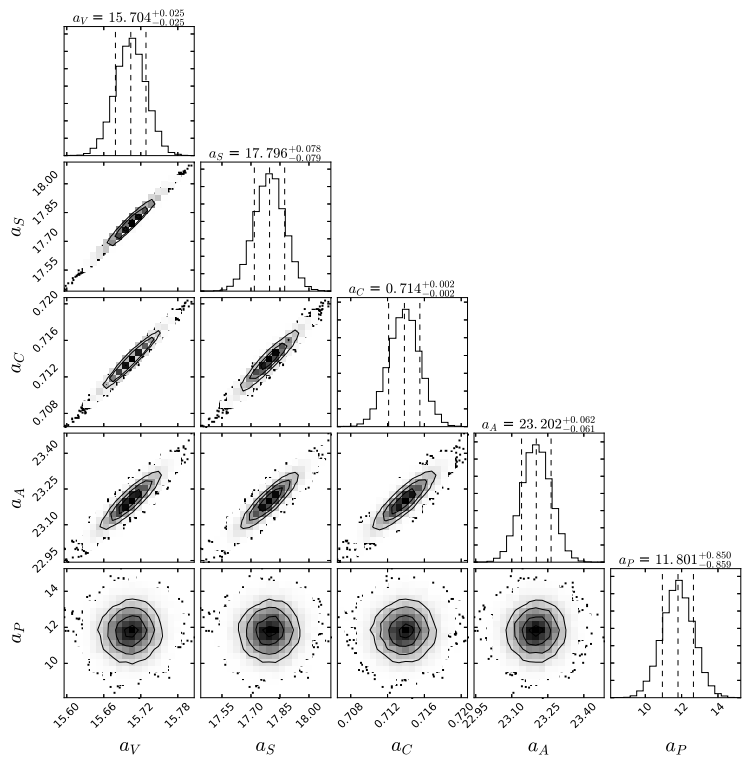}
}
\caption{Corner plot of the exact likelihood surface in vicinity of the maximum. The diagonal plots show the parameters distributions. The central dashed line corresponds to the mean of the distribution, while the side ones delimit the 1$\sigma$ interval. The off-diagonal plots show the correlation between the two parameters. See text for details.}
\label{fig:analytical_corner}
\end{figure}

As a start, we explore the likelihood surface associated to our penalty $\chi^2$ function using  the No-U-Turn Sampler (NUTS)~\cite{Homan:2014:NSA:2627435.2638586}, an efficient Markov chain Monte Carlo (MCMC) sampler~\cite{neal1993probabilistic}.
In  Fig.\ref{fig:analytical_corner}, we present the marginalised likelihood in vicinity of the maximum for individuals parameters on the diagonal part and the contour plots for pairs of parameters off the diagonal. The best estimation of LDM parameters with error bars,extracted as the 68\% percentile of each parameter distribution, are given on top of each column.

The marginalisation of the likelihood~\cite{mcd15} provides us with a very useful insight on the behaviour of the likelihood surface.  Given its exponential nature, the MCMC algorithm is the most suitable technique to perform the required multidimensional integrals leading to the marginalisation.
From the  resulting  distributions of the parameters  $\mathbf{p}$, we can extract information about correlations. In particular, we have been able to directly extract the covariance matrix without performing any derivative in parameter space~\cite{bar89}. In Tab. \ref{tab:corr}, we provide the correlation matrix extracted from MCMC. We observe that the results are in perfect agreement with the correlations obtained using Non-Parametric Bootstrap~\cite{pas18b}, which is also a Monte-Carlo sampling based on the log-likelihood function.
For this particular model, we observe a strong correlation among the parameters $a_v,a_s,a_a,a_c$ that also reflects on the cigar-like shape of the marginalised likelihood shown in Fig.\ref{fig:analytical_corner}. The pairing term is not correlated to the others, thus the marginal likelihood has a spheroid shape.

\begin{table}
\begin{center}
\begin{tabular}{c|ccccc}
\hline
\hline
&$a_v$ & $a_s$ &$a_c$ & $a_a$ & $a_p$\\
\hline 
 $a_v$& 1 &        & & & \\
$a_s$  &0.993 &       1 &       & & \\
$a_c$  &0.984 &       0.962 &       1 &      &\\
$a_a$  &0.917 &      0.901 &       0.884 &        1&\\
$a_p$&0.038 & 0.037 &  0.040 &  0.038&   1\\  
\hline
\hline
\end{tabular}
\caption{Correlation matrix for LDM parameters obtained from MCMC sampling.}
\label{tab:corr}
\end{center}
\end{table}

MCMC is a valuable tool to explore the surface of the unimodal  likelihood, but in the case of a multimodal one, it may not be the most efficient method. A possible way to solve the issue is to use the GPE method to emulate the likelihood surface together with  Expected Improvement (EI) criterion \cite{jonesEfficientGlobalOptimization1998} to focus the exploration around the maxima.

The EI criterion helps us to choose  iteratively the points  in parameter space one should explore in order to efficiently find a global optimum of a function. It is an optimisation technique which, after a large enough number of iterations, will converge to the global optimum, never getting stuck into a local optimum.
It uses both the GPE mean and GPE confidence intervals, in a tradeoff between exploitation and exploration. The first means sampling areas of likely improvement near the current optimum predicted by the mean, the second means sampling areas of high uncertainty where the confidence intervals are large. It employs an acquisition function, whose maximum tells us where next in parameter space to run our model. GPE is then performed again with this new point, and this process is repeated until  convergence  is reached according to a given convergence criteria.

For the sake of illustration, we fix three parameters, $a_c$, $a_a$, $a_p$, of LDM at values obtained from MCMC and emulate the log-likelihood surface, $i.e.$ the $\chi^2$, as a function of the remaining two parameters $a_v$, $a_s$.


\begin{figure}[h!]
\centerline{
\includegraphics[width=0.48\textwidth]{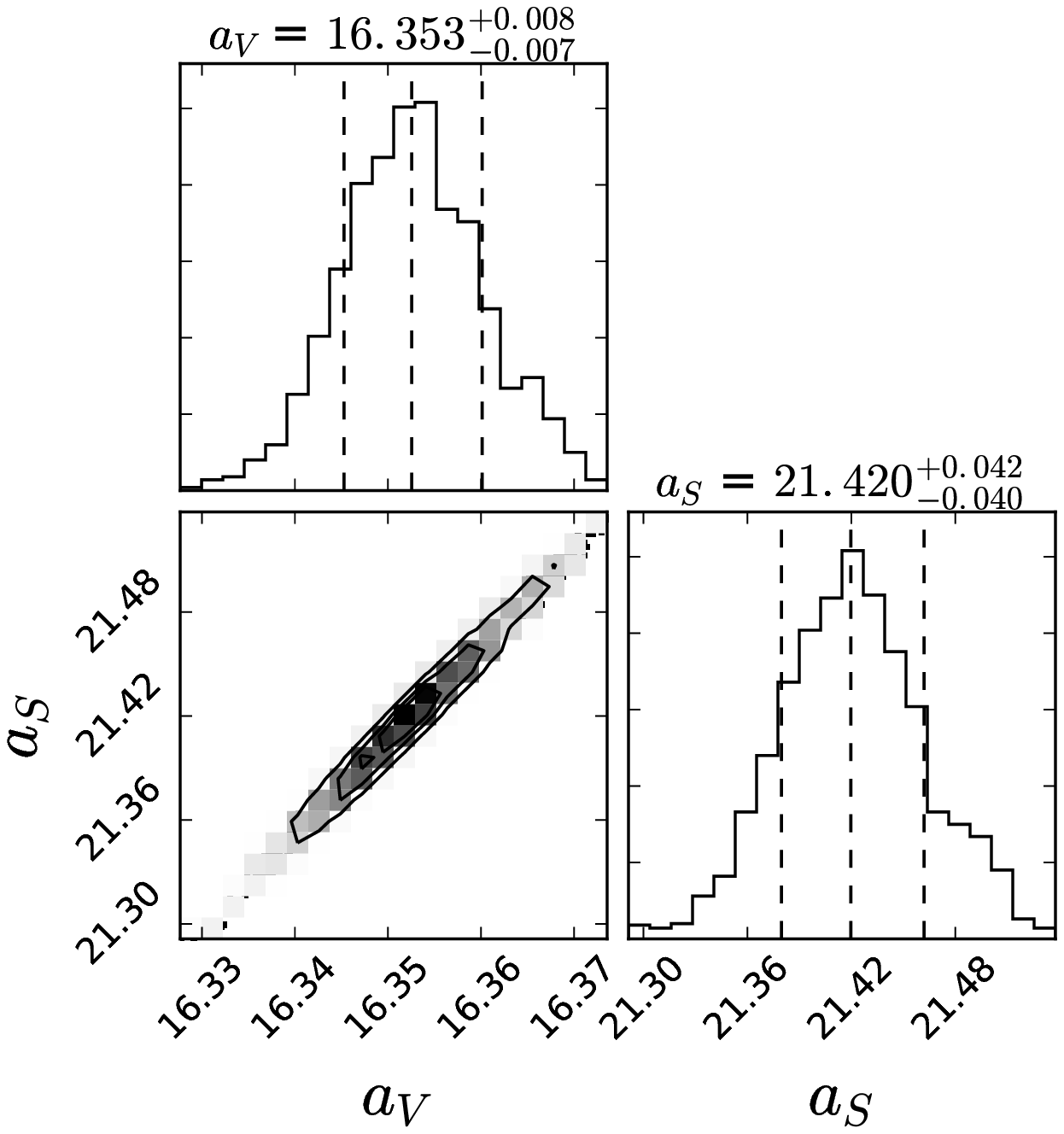}
\includegraphics[width=0.48\textwidth]{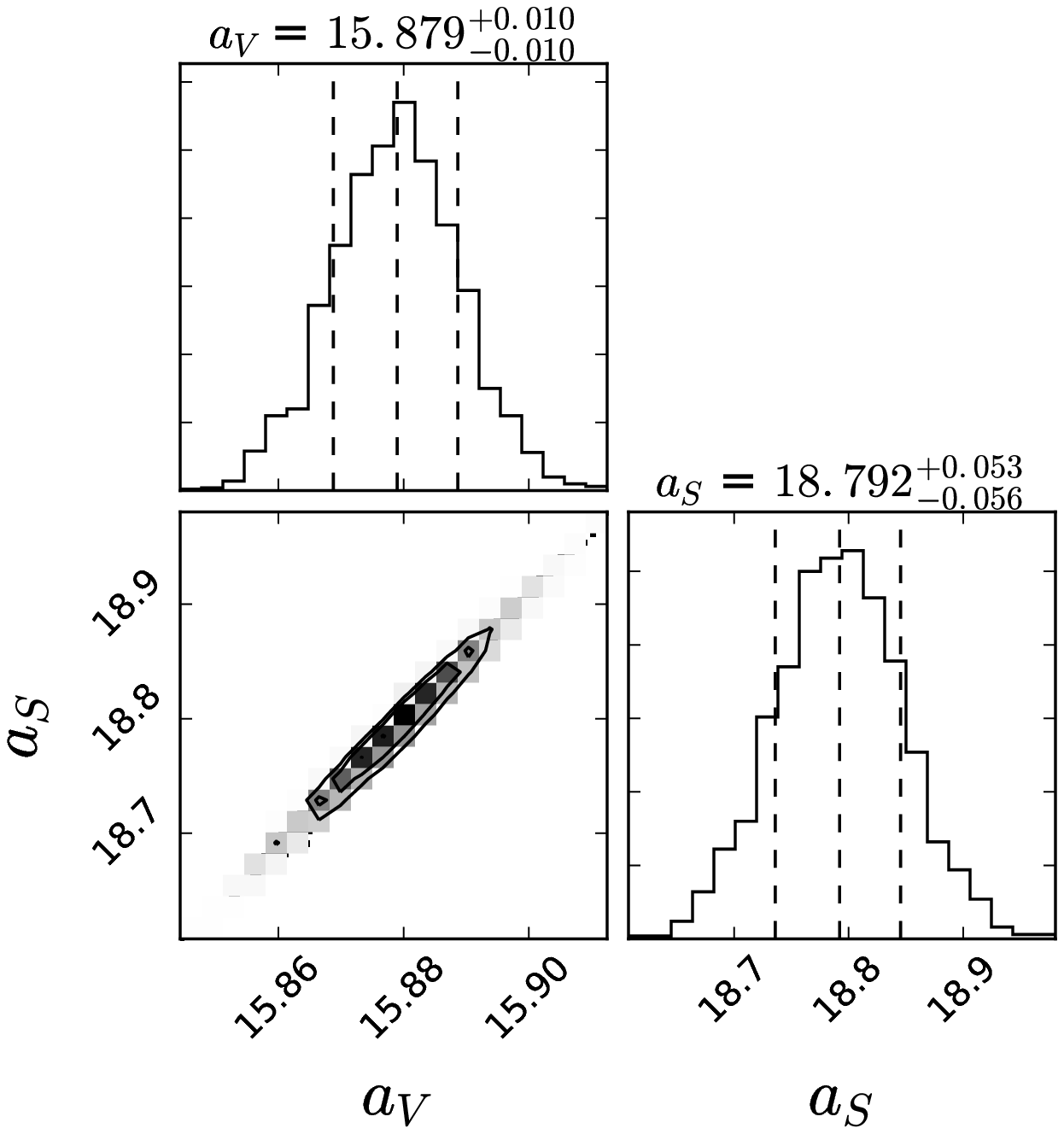}
}
\centerline{
\includegraphics[width=0.48\textwidth]{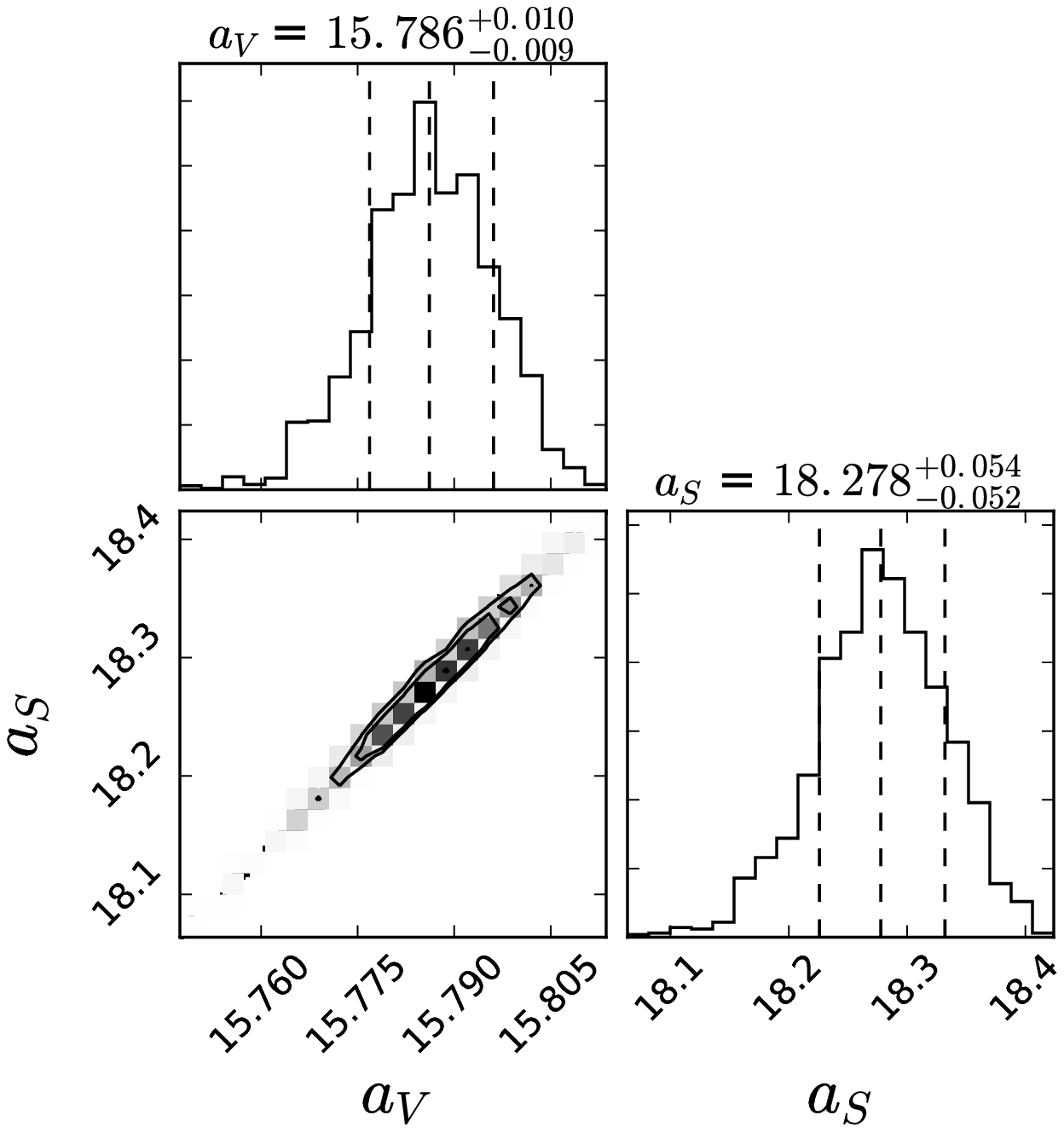}
\includegraphics[width=0.48\textwidth]{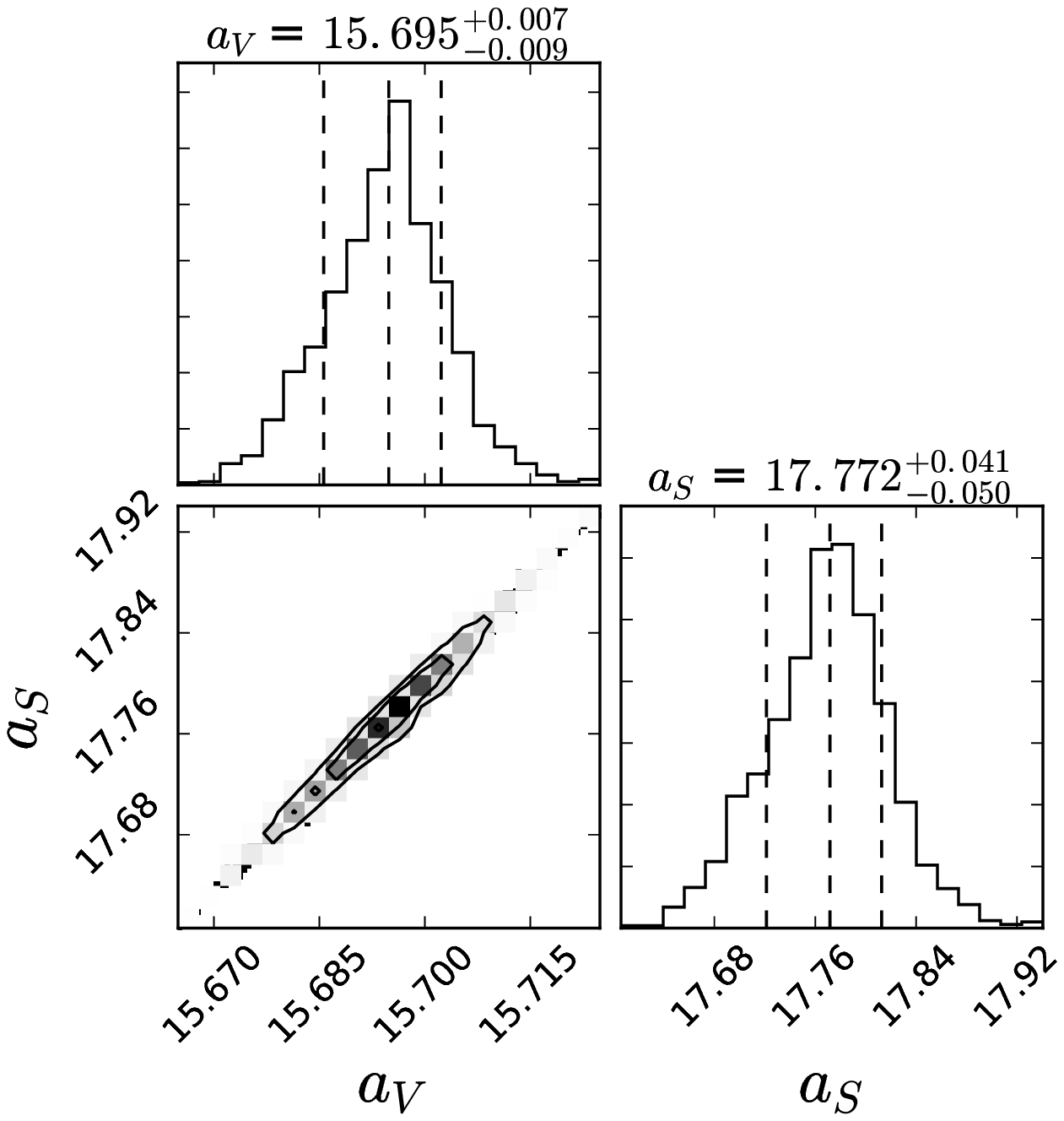}
}
\caption{Corner plots of the likelihood obtained from the emulated $\chi^2$ surface for the $a_v$ and $a_s$ parameters, after the first (top left), 11 (top right), 21 (bottom left) and 51 (bottom right) iterations of EI. See text for details.}
\label{fig:gpe_corner}
\end{figure}

We design the initial data set by taking 60 points for each dimension to ensure a good quality emulation from the outset of the optimisation.  We impose a uniform prior distribution on parameter space, by restricting to a fairly large interval $a_{v(s)}\in[0,30]$ MeV. The data-set is selected using  a standard space-filling method~\cite{mckayComparisonThreeMethods1979}. 
After running GPE, we feed the resulting likelihood surface to the MCMC sampler.
The result is illustrated in  Fig.~\ref{fig:gpe_corner}. In the top left panel, we show the result obtained using the initial data-set, $i.e.$ without any EI. We notice that the means  of the $a_v$ and $a_s$  distribution  are quite different from the expected optimal values (see Fig.\ref{fig:analytical_corner}), that we attribute to an incorrect emulation of the likelihood surface.

We now apply an iterative procedure guided by EI: in the top right panel we add extra 11 points and then we perform a new MCMC run. In this case the volume parameter gets extremely close to its optimal value and errors.

The iterative EI procedure was then carried out in the bottom left panel (21 iterations) and bottom right (51 iterations) until the mean value of the distributions of the parameters $a_v,a_s$ fall within the error bars provided by the complete MCMC sampling.
As expected, in the limit of large number of points,  GPE will provide a very good approximation of the exact surface.


\section{Conclusions}\label{sec:conc}

In this article, we have explained how advanced statistical methods may be a very useful tool to estimate parameter in theoretical nuclear models. 
By using a simple Liquid Drop Model, we have discussed the use of MCMC method to visualise multidimensional likelihood surfaces and how to extract information concerning the covariance matrix without performing numerical derivatives in parameter space. This is a very interesting method, since contrary to the standard Hessian method~\cite{roca2015covariance}, we do not need to assume a parabolic shape of the surface around the optimum.

Since MCMC is likely not adequate to deal with multimodal likelihood surfaces, we have also investigated the use of GPE combined with the EI to identify in an efficient way the position of the optimum in parameter space and then use MCMC to obtain informations on the parameter distributions.
By combining these two methods, we now plan to perform a more complete study using the likelihood surface of a Skyrme functional~\cite{per04}. 



\section*{Acknowledgments}


The work  is supported  by the UK Science and Technology Facilities Council under Grants No. ST/L005727 and ST/M006433. 

\bibliographystyle{polonica}
\bibliography{biblio}

\end{document}